Also available at www.brill.nl.beh

# Group movement decisions in capuchin monkeys: the utility of an experimental study and a mathematical model to explore the relationship between individual and collective behaviours

Short title: Group movement decision in capuchin monkeys

H. Meunier[1,2], J.-B. Leca[1], J.-L. Deneubourg[2] & O. Petit[1]

[1] Ethologie des Primates, IPHC, Département Ecologie Physiologie et Ethologie, UMR 7178 CNRS-ULP, 67087 Strasbourg Cedex 2, France

[2] Service d'Ecologie Sociale, Université Libre de Bruxelles, Belgium

[1] Corresponding author's e-mail address: helene.meunier@c-strasbourg.fr

**Summary**

In primate groups, collective movements are typically described as processes dependent on leadership mechanisms. However, in some species, decision-making includes negotiations and distributed leadership. These facts suggest that simple underlying processes may explain certain decision mechanisms during collective movements. To study such processes, we have designed experiments on white-faced capuchin monkeys (*Cebus capucinus*) during which we provoked collective movements involving a binary choice. These experiments enabled us to analyse the spatial decisions of individuals in the group. We found that the underlying process includes anonymous mimetism, which means that each individual may influence all members of the group. To support this result, we created a mathematical model issued from our experimental data. A totally anonymous model does not fit perfectly with our experimental distribution. A more individualised model, which takes into account the specific behaviour of social peripheral individuals, revealed the validity of the mimetism hypothesis. Even though white-faced capuchins have complex cognitive abilities, a coexistence of anonymous and social mechanisms appears to influence their choice of direction during collective movements. The present approach may offer vital insights into the relationships between individual behaviours and their emergent collective acts.

## Introduction

In group-living animals, a wide range of behaviours like resting, foraging or moving may be performed collectively. The functions of such groupings are diverse: antipredation (van Schaik, 1983; Sterck *et al.*, 1997; Isbell, 1994), foraging benefits (Wrangham, 1980; Terborgh, 1983) and energy saving (Weimerskirch *et al.*, 2001). In a social group, animals have different motivations and have to compromise between their own interests and the costs of a collective choice which could differ from their own needs. In the case of group movement, if all members choose different directions, the group will split and its members may lose many of the advantages of group living (Krause & Ruxton, 2002). Observational and empirical evidence shows that animal groups move across the landscape quite cohesively (Stewart & Harcourt, 1994; Boinski & Campbell, 1995; Boinski, 1996, 2000; Byrne, 2000; Parrish *et al.*, 2002; Conradt & Roper, 2003), which strongly suggests that a collective decision has been taken. Thus it could be assumed that individual decisions lead to a common decision, allowing the group to remain cohesive (Conradt & Roper, 2005).

Classically, collective movements in primate groups are described as processes dependent on leadership mechanisms where a single individual initiates a group movement and is followed by other individuals (Boinski & Garber, 2000). Mountain gorillas (*Gorilla gorilla beringei*) are the best known example of such leadership concentrated on a single individual (Schaller, 1963). However, it has been suggested that instead of one individual being responsible for a decision, a division of roles among an initiator and other decision-making individuals may exist (Byrne, 2000). In Drakensberg mountain baboons (*Papio ursinus*), Byrne *et al.* (1990) have described several scenarios ranging from a single individual who initiates and determines the direction and the departure time, to the case where the initiator seems to be ineffective alone and needs the adhesion of a first follower to influence the group. In white-faced capuchins (*Cebus capucinus*), Leca *et al.* (2003) showed that: (1) several individuals, not necessarily the most dominant ones, can initiate movements, and (2) the spatial and temporal distribution of the group affects the probability that other group members respond positively to an initiation.

These observations lead us to consider alternative mechanisms which may help us to understand exactly how a collective decision is reached. Mimetic interactions between group members could play a key role in collective movements. Mimetic behaviour, where animals act like their conspecifics, is widespread in animal societies and is an example of positive behavioural feedback (Sumpter, 2006). Distributing the team within the environment and introducing positive feedback among animals allows amplification of the decision taken by a few individuals. Through competition among different amplifications, all individuals reach a consensus decision and maintain group cohesion (Deneubourg & Goss, 1989; Bonabeau *et al.*, 1997; Detrain *et al.*, 1999; Camazine *et al.*, 2001; Deneubourg *et al.*, 2002; Jeanson *et al.*, 2004; Couzin *et al.*, 2005; Amé *et al.*, 2006). Such self-organized processes allow groups to carry out collective actions in various environments without any lead, external control or central coordination. They are used by many species living in large societies (Conradt & Roper, 2005) with low individual levels of cognition but also by vertebrates, including primates, living in small or large groups (Parrish & Edelstein-Keshet, 1999; Hemelrijk, 2002; Couzin & Krause, 2003).

Our objective was to test whether anonymous or social processes govern the choice of a given direction for collective movements in a white-faced capuchin monkey group. In order to assess the mechanisms of the collective movements and their dynamics, we have designed experiments where we provoked collective movements to explore the process in a purest form, which is practically impossible to achieve in a natural environment. Based on these experiments, we analysed spatial decisions within the group. Our hypothesis is that the nature of the underlying processes concerns mimetism (acting as a catalyst for the collective decision) combined with individualities. To validate this hypothesis, we use a mathematical model to explore the relation between individual behaviour and collective phenomena.

## Material and methods

### *Subjects and environment*

The group of white-faced capuchins was established in 1989 at the Louis Pasteur University Primate Centre, Strasbourg, France. During our study, the group contained 13 individuals of three separate lineages: Five males (aged 2, 4, 8, 9 and 20 years or more) and eight females (2, 2, 2, 7, 8, 9 years old, and two individuals aged 20 years or more).

The group was kept in a one-acre outdoor enclosure with natural vegetation and uneven ground with free access to an indoor shelter. Commercial primate pellets and water were available *ad libitum*. Fresh fruits and vegetables were provided once a week but not during testing.

### *Observation procedure*

Observations took place between 0900 and 1200 hours and between 1400 and 1800 hours from May to August 2001. Three observers collected data with video and tape recorders and communicated using walkie-talkies.

The first phase of the experiment consisted in training the capuchins to move to the sound of a whistle. The sound of the whistle was gradually associated with the subsequent presentation of food located further and further away from the starting point where the sound was emitted. At the end of the training period, the blast of the whistle was perceived as a food-anticipatory signal leading to the possible presence of food in a remote location. The second phase of the experiment consisted of 108 tests during which the capuchins had the opportunity to choose between two opposite directions leading to two distinct areas in the park. In non-experimental baseline context, animals spontaneously used a particular zone of the enclosure for social and resting activities. This zone is referred to as the "departure zone". The two areas to choose from are natural foraging areas situated 60 meters away from the departure zone. A manger was placed in each of the areas but was not visible from the departure zone.

During each test, only one randomly selected manger contained figs. The other one was left empty. The same manger could not be filled more than three times successively to prevent learning of reward position. Each manger was 2 meters long thus all monkeys could feed simultaneously from the same manger. When all animals were grouped in the departure zone, the whistle was sounded, and the mangers were opened. The choice for left or right manger was made from the very beginning. The initial direction taken by the animals from the departure zone was systematically maintained until the chosen manger was reached. The direction taken by each monkey was recorded for each test and coded by an "L" when animals chose to go left and by an "R" when they chose to go right.

***Social relationships***

To establish the dominance hierarchy, we ranked individuals over 1 year of age in a matrix according to the direction of avoidances and unidirectional aggressions. We used data from two contexts: (1) spontaneous events and (2) drinking competition around a single source of orange juice (three series of nine 2-h tests). We carried out hierarchical rank order analysis using Matman, (de Vries *et al.*, 1993). We verified the linearity of the dominance hierarchy, $h'$=0.91 ($P$<0.001; de Vries, 1995). The dominance scores ranged from one for the most dominant individual to thirteen for the most subordinate one.

Affiliation was quantified by the frequency of body contacts among all identified group members, recorded by using instantaneous sampling every 5 minutes (Altmann, 1974).

We collected 728 scans, but not during fruit provisioning. The affiliation score within each dyad was assessed by the number of scans during which the two partners were in body contact.

### *Data analysis*

For each individual, we obtained a total of spatial association, defined as the total number of group members that chose the same direction as this individual across all tests. Several matrices were built to analyse the effect of socio-demographic variables on spatial association. We firstly reported in a symmetrical matrix when one animal chose the same direction as another one, i.e. the frequency of spatial association choice for each dyad. We also reported the affiliation scores in a symmetrical matrix. Finally, we assessed the degree of closeness in maternal kin relationships by distinguishing three types of dyads: non-kin, far-kin (siblings, half siblings, grandmother-grandchildren, aunt-nephew/niece), and close-kin (mother-offspring) dyads. We implemented the degree of kinship in the three types of kin dyads in a matrix. Matrix correlations were tested by using Matman (de Vries *et al.*, 1993). We set the number of automatic permutations of matrices at 10000 and used Pearson's correlation coefficient.

Non-parametric statistical tests used were the Kolmogorov-Smirnov test, Kruskal-Wallis test, Spearman rank correlation test and the chi-square test (Siegel & Castellan, 1988). All tests were two-tailed and the significance level was fixed at 0.01.

For decision analysis, only 64 tests were used because of missing data on individual direction choice in the other tests.

## Results

### *Description of the experiment*

At the sound of the whistle, individuals started to move, one after one. The mean starting time was 7.5 ±0.6 sec for the first individual and 170.2 ±16.5 sec for the whole group to have moved. When leaving the departure zone, the monkeys chose between the two directions leading to the two mangers one after one. All decisions were made prior to vocalizations from travelling individuals who had already arrived at the manger. Before reaching the manger corresponding to the initial direction chosen, no animal was observed returning to the departure zone or going towards the opposite manger.

***Independency of the tests***

The localisation of the reward in a given test (in the left or the right manger) did not affect the choice of the animals in the subsequent test: the majority of the group chose 50 times the manger filled previously in the following test (out of 107 subsequent choices). This result suggests that reward-reinforcement bias should not be invoked in our study.

***Individualities and relationships***

During collective movements, capuchins were not significantly associated according to sex (Kruskal-Wallis test: $\chi^2$=3.45, $N_{male-male}$ =10, $N_{female-female}$ =28, $N_{male-female}$=40, p=0.178) or dominance rank (Spearman rank correlation test: $r_s$ = -0.110; N = 13; p = 0.721 ) regarding direction choice. Moreover, matrix correlations revealed that kinship (Pearson's correlation coefficient r = -0.027; p = 0.506) and affiliation (Pearson's correlation coefficient r = 0.236; p = 0.054) did not influence the spatial associations of capuchins.

Twelve out of thirteen individuals started to move first at least once. First position frequency was not significantly correlated with dominance rank (Spearman rank correlation test: $r_s$=0.397, N=13, p=0.180).

***Decision Analysis***

We tested whether the monkeys followed their conspecifics or if they chose a direction independently of other group members. Of the 64 tests, the total number of choices was 832 (13 individuals x 64 tests), and the proportions were respectively 0.4 for the L side and 0.6 for the R side. This asymmetry suggests a weak preference for the right side and was taken into account in the subsequent analysis.

The frequency of tests where $i$ individuals ($i$ = 0,…,13) chose the same direction was measured. Assuming that the individuals selected their side independently with the probabilities $P_L = 0.4$ and $P_R = 0.6$, the theoretical distribution of the tests as a function of the number of individuals choosing the L or R is a binomial distribution (Figure 1a). However, our experimental distribution is bimodal, a characteristic of a collective choice (Camazine *et al.*, 2001), and differs from the binomial theoretical one (Chi-square test: $\chi^2$=37.067, df=13, p=0.004, Figure 1). To test the process of collective choice and inter-individual influence, we made a further analysis by considering the directions taken by the n first individuals (n=2,…,13). If the n first individuals chose their direction independently, the probability ($P_n$) that they chose L (R) is $P_L^n$ ($P_R^n$). We compared the number of tests where at least n first individuals have taken the same direction to the theoretical ones (64 x ($P_L^n + P_R^n$)). The experimental distribution was statistically different and higher than the theoretical one and the maximum difference was observed for the six first individuals (Kolmogorov-Smirnov two-sample test: D=0.67, N1=N2=13, p=0.015). The ratio between $P_n/P_{n-1}$ increases with n, showing that the larger the number of individuals having chosen one side, the higher the probability is that the following individuals will also choose this side. The experimental probabilities are always higher than the theoretical ones.

This result shows a correlation between the choices of the individuals, which probably results from a mimetic effect: each individual seems to be both influenced by the choice of the others and have a tendency to follow the direction taken by the previous one(s).

***Independent Individuals***

In this experimental group, individuals have two different profiles. For each individual, we compare the distribution of the number of tests in which it has taken the same direction as 0, 1, 2, …, 12 of its groupmates, with a distribution calculated from the mean of the corresponding values of all the other individuals (number of times that 0,…,12 individuals take the same direction). Ten individuals have similar profiles (their individual distributions do not show significant differences to the calculated one), and they present a tendency to follow other group members (‘dependent’ individuals). The three other individuals (subordinate females) behave differently from the other groupmates (chi-square test: $\chi^2$=44.26, 85.66 and 86.39 for those three animals, df=12, p=0.010). These ‘independent’ individuals tended to choose a side independently of others and thereby move either to the side mainly chosen by group members or to the other side.

***Anonymous model: all individuals identical***

Here we test the mimetic hypothesis to explain the collective mechanisms involved in the group choice. In this model, all individuals (*N*) are identical. Any individual has the probabilities $P_L$ to go L and $P_R$ to go R that depend on its intrinsic preference to choose the left ($\alpha_L$) or right ($\alpha_R$) side and on the decision of the previous individuals.

The first decision is the choice of the first individual to take the left or the right path. In this case, $P_R = \alpha_R$ and $P_L = \alpha_L$.

Similarly we simulate the decision of the second,…, thirteen individual. To take into account the mimetic behaviour, $P_L$ ($P_R$) must increase with $N_L$ ($N_R$) and decrease with $N_R$ ($N_L$). $N_L$ and $N_R$ are the number of individuals having chosen the left and right side and $N_L + N_R$ are the individuals having moved before the individual. The simplest form of $P_R$ or $P_L$ is:

$$P_L = \frac{\alpha_L + \beta N_L}{1 + \beta(N_L + N_R)} \qquad (1,a)$$

$$P_R = \frac{\alpha_R + \beta N_R}{1 + \beta(N_L + N_R)} \qquad (1,b)$$

$$P_R + P_L = 1 \qquad \alpha_R + \alpha_L = 1$$

$\beta$ is the mimetic coefficient that takes account of the influence of the individual having previously moved and decided. If $\beta = 0$, there is no mimetic behaviour and the individuals act independently of each other. The greater the value of $\beta$ the greater the mimetism.

In order to establish the main factors causing the fluctuations of the experimental results, we used Monte Carlo simulations. In such a numerical simulation, the random aspects of the process are thus automatically incorporated.

We can summarize the different steps as follows:

- Initial condition: At the beginning of the simulation, all the individuals are at rest, ($N_R = N_L = 0$).
- Decision process: The decisions of the N individuals are tested.

To determine the choice of an individual, the value of a random number is compared to $P_L$ (equation 1,a) , depending on the choice of the previous individuals. For each monkey, the random number is drawn from a uniform distribution between 0 and 1. If its value is less than or equal to $P_L$ the monkey chooses the left side and if the number is greater than $P_L$, it chooses the right side. The delay between two departures was not considered.

The simulations are run 1000 times for the $N$ animals and we calculate the distribution of the simulations as a function of the number of individuals having chosen the left side.

We let the $\beta$ mimetic parameter vary and run simulations for frequency of choices for the same side. The case $\beta = 0$ corresponds to individuals acting independently and the distribution is binomial (Figure 1a). For very low values of $\beta$, the model also exhibits unimodal distributions. Three simulations with three different $\beta \geq 1$ are presented in Figure 2. A strong bimodal distribution is only observed for large values of $\beta$ ($\beta \geq 3$) but the fraction of the simulations characterized by most of the individuals having chosen the same side is much greater than the corresponding experimental ones (Figure 2b & 2c).

***Individualised Model***

We modified the previous model by individualising the process of directional choice. In this model, each individual is characterized by a specific function $P_{Li}$ and $P_{Ri}$.

We assume two categories of individuals: (1) mimetic individuals having the same behaviour as in the previous models, (2) independent subjects that are not influenced by and reciprocally do not influence the other individuals (no mimetic process for direction choice process). The behaviour of these independent subjects corresponds to the behaviour of the three experimental peripheral females.

$$P_{Li} = \frac{\alpha_{Li} + \beta_i \sum_{j=1}^{N} \delta_{ij} L_j}{1 + \beta_i \sum_{j=1}^{N} \delta_{ij} (R_j + L_j)} \qquad (2,a)$$

$$P_{Ri} = \frac{\alpha_{Ri} + \beta_i \sum_{j=1}^{N} \delta_{ij} R_j}{1 + \beta_i \sum_{j=1}^{N} \delta_{ij} (R_j + L_j)} \qquad (2,b)$$

$$P_{Ri} + P_{Li} = 1 \qquad \alpha_{Ri} + \alpha_{Li} = 1$$

$\beta_i = 0$ corresponds to an independent individual not influenced by the others. $\delta_{ij} = 1$ if the individual *j* could influence the individual *i*. $\delta_{ij} = 0$ if the individual j could not influence the individual *i*. $L_j = 1$ and $R_j = 0$ ($L_j = 0$ and $R_j = 1$ ) if the individual *j* has previously chosen the left side (right side). $L_j = 0$ and $R_j = 0$ if the individual *j* has not yet moved. We can note that if all the dependent individuals present the same mimetic behaviour ($\beta_1 = ... = \beta_N$, $\delta_{ij} = 1$), equations 2 are equal to equations 1. We have performed simulations based on our experimental observations with three independent individuals and ten mimetic individuals.

We plot the same distribution as for the anonymous model: the distribution of the simulations as a function of the number of individuals having chosen the left side for three mimetic coefficients (Figure 3). The results taking into account the two types of individuals were similar to the experimental distributions: the obtained distribution was bimodal and the maximum values of the two modes were very close to those of the experimental distribution.

## Discussion

In this study we have demonstrated that most capuchin monkeys tend to follow the travel route previously taken by their groupmates when given a binary choice, but that a minority of individuals consistently decides their route independently from their groupmates behaviour.

### ***Experimental spatial decisions***

The experimental distributions of the number of individuals moving in the same direction are bimodal, which is the signature of a phenomenon involving interactions between individuals (Camazine *et al.* 2001). At departure time, all group members were grouped together in the departure area, where each capuchin had the opportunity to observe the behaviour of other group members. One can propose facilitation of the response in this case. In such a process, the presence of a group member performing an act already within the observer's repertoire increases the probability of the observer reproducing that act (Byrne 1994). In our study, the facilitated response was the directional choice. The departure of one or more individuals to a given side can draw the attention of the others. This would involve a facilitation of the moving of individuals still present in the departure zone, in the same direction. This phenomenon could also be interpreted from the perspective of group cohesion in relation to predation risk. As more individuals have left the departure area, the chance for a capuchin to be left alone in this area increases. In many cases, such situation is potentially hazardous in terms of predation for an individual living in a wild primate group. This may account for a natural tendency to grouped departures in travelling primates.

We found that this collective pattern is not a consequence of demographic and social relationships between group members. Indeed, during movements, capuchins do not follow their groupmates according to their sex, dominance rank, kin or affiliative relationships. It seems that this phenomenon is anonymous from a social point of view, i.e. does not depend on individualities or social relationships. Moreover, no vocalization was emitted by capuchins en route for the arrival area. The environmental differences between our captive conditions and the wild may partly explain the absence of trill vocalizations in this specific experimental context, where the locations of the mangers were well known by the monkeys. Conversely, trills have been emitted by the same group members in the context of spontaneous moves where only the initiator took the lead (Leca *et al.*, 2003). In the wild, where food sources are more dispersed and less limited, white-faced capuchins also use trill vocalizations to coordination troop movements (Boinski & Campbell, 1995).

Distributions observed experimentally also show that the group was split into two subgroups. Subgroup sizes were asymmetrical with generally one sub-group of ten or eleven individuals and the other of two or three. This supports the hypothesis of a collective movement and discredits the hypothesis proposing that all individuals should behave independently of each other. In the case of truly independent individuals, the distribution obtained would have been of a binomial type.

This result however gives no indication of the underlying mechanisms of the asymmetrical division of the group. A finer analysis of spatial decisions whatever the identity of the individuals reveals that the greater the number of individuals choosing one side, the greater the probability that the other remaining individuals will choose the same side. This behaviour has been described by the concept of contagion (Thorpe 1963), which is realised through mimetic processes. However our results show that not all group individuals are involved in the chain of contagion or display mimetic behaviour.

***Modelling***

To validate our hypothesis of mimetic processes, it was important to draw up models (Deneubourg & Goss 1989; Camazine *et al.* 2001; Sumpter 2006). This approach enables the simulation of a large number of events starting from a model. We could implement mechanisms deduced from our decision analysis. In the anonymous model, all individuals were considered as identical entities. Regarding non-human primates, it is clear that the hypothesis of equal individuals is coarse (Stevenson-Hinde 1983) but this simple formulation remains an important first step. It is a traditional approach in modelling which shows the role of anonymous processes in the phenomenon studied (Camazine *et al.* 2001). The response of the first model supports our hypothesis of an anonymous process but cannot account for all the observations. The distributions obtained from this simple model are bimodal like experimental ones and confirm the hypothesis of subjacent mimetism for the movements, making it possible to seize the logic behind the phenomenon. However, the quantitative responses of the model are not satisfying enough because they are too pronounced.

In the individualised model, we retain the basic structure (the same mechanisms are implemented) but we take some independent individuals into consideration. Concerning directional decisions, three individuals present differences in their behaviour and are qualified as independent. It can be mentioned that these three individuals are subordinate and frequently peripheral individuals (socially and spatially). Such peripheral positions have been described in groups of wild white-faced capuchins (Perry 1996, 1997, 1998; Rose 1994) and other primate species during collective movements (baboons: Rhine 1975; Rhine & Westlund 1981; lemurs: Kappeler 2000). This could be explained by the fact that subordinate individuals suffer the most feeding competition (Whiten, 1983; Saito, 1996; Sterck *et al.*, 1997; Wittig & Boesch, 2003 ) and may adopt particular strategies, such as moving alone in the outskirt of the group, which could be beneficial when food competition is too important. Moreover, spatial separation may also enhance individual opportunities for feeding innovation (e.g., Leca et al., in press). The independent individuals are taken into account in the model by assigning them a mimetic coefficient of zero. With inclusion of such parameters, the model produces distributions very similar to the experimental one and reproduces the majority of our collective movements.

These results enable us to conclude the coexistence of different individual profiles (dependent or independent) in collective movements of white-faced capuchin monkeys. We are fully aware that our results are limited to one social group and supplementary data from additional groups are required for our conclusions to be generalized. Nevertheless, our approach, based on experimental testing and modelling of the experimental results, allows us to reveal for the first time that part of primates collective behaviour could be anonymous. This goes against some generally accepted ideas on primate societies and the mechanisms underlying collective phenomena in such groups. Because of the developed cognitive capacities and complex social systems of primates, collective phenomena are generally perceived as being based on multiple interactions, communication and negotiation and are therefore difficult to model. Our approach was relatively simple. It models catalytic mechanisms and basic mimetic process and allows us to reproduce the collective choice of our group. To draw more general conclusions concerning group movement decision process in primates, this type of study should to be conduted in other groups of white-faced capuchins and in other species, and ultimately our results should be verified in more natural conditions.

**Acknowledgements**

We are grateful to the Louis Pasteur University Primate Centre, Strasbourg, France and to P. Uhlrich for technical support for the experiments. J.L.D. is research associate from the Belgian national Funds for Scientific Research. We also thank the two anonymous reviewers for insightful comments on the manuscript.

**Figures**

Figure 1

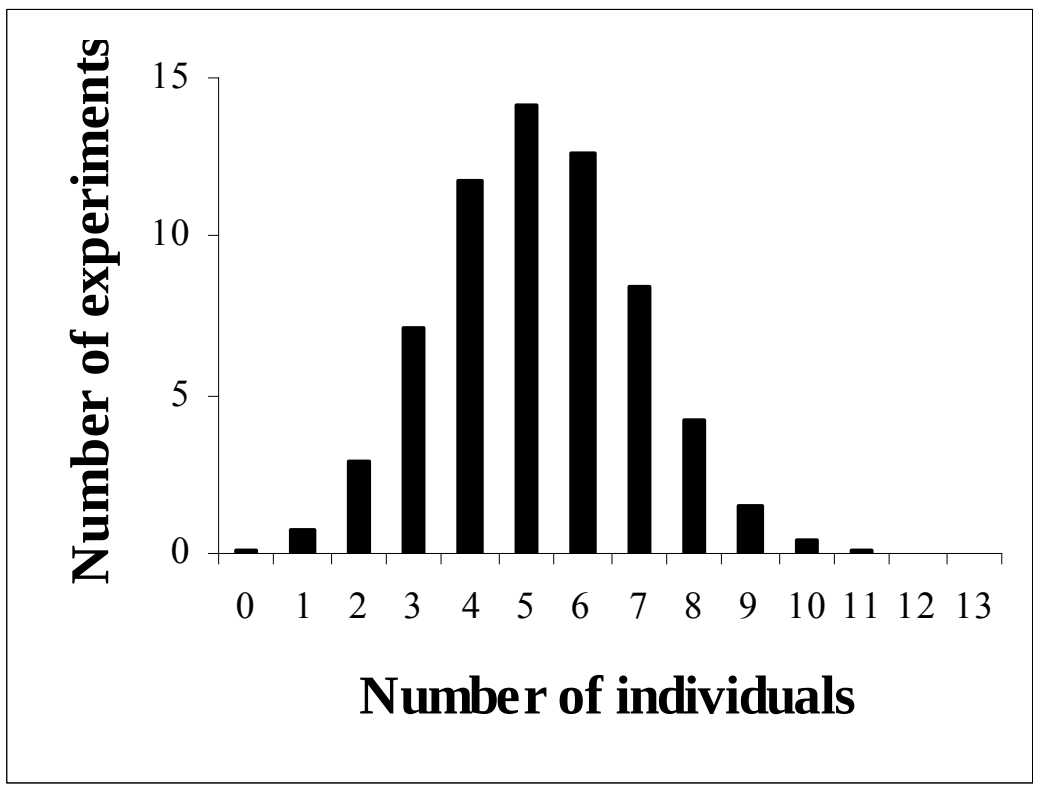

a

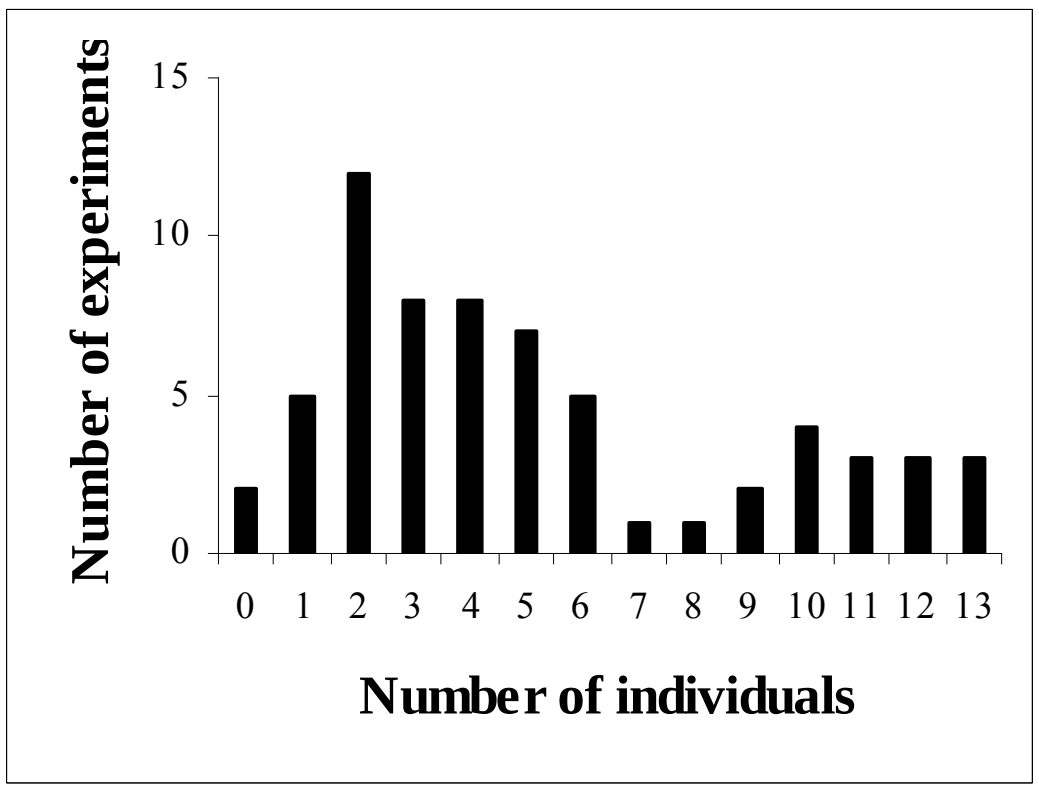

b

Figure 2

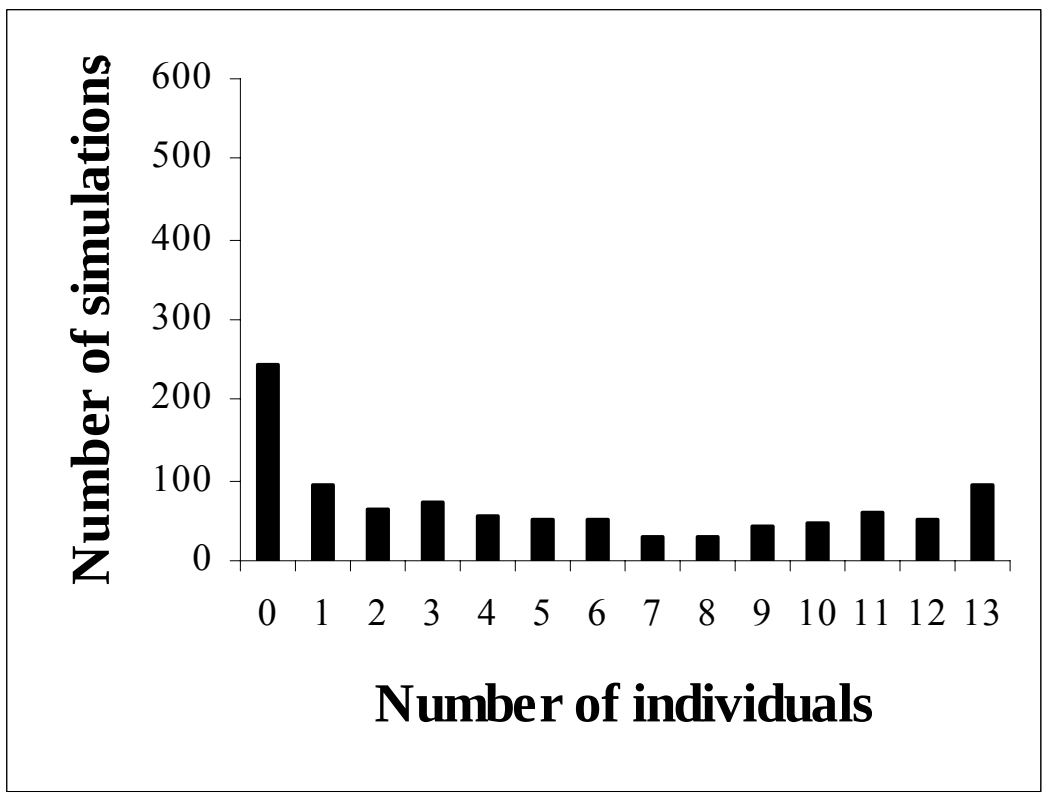

Number of simulations
600
500
400
300
200
100
0
0 1 2 3 4 5 6 7 8 9 10 11 12 13
Number of individuals

a

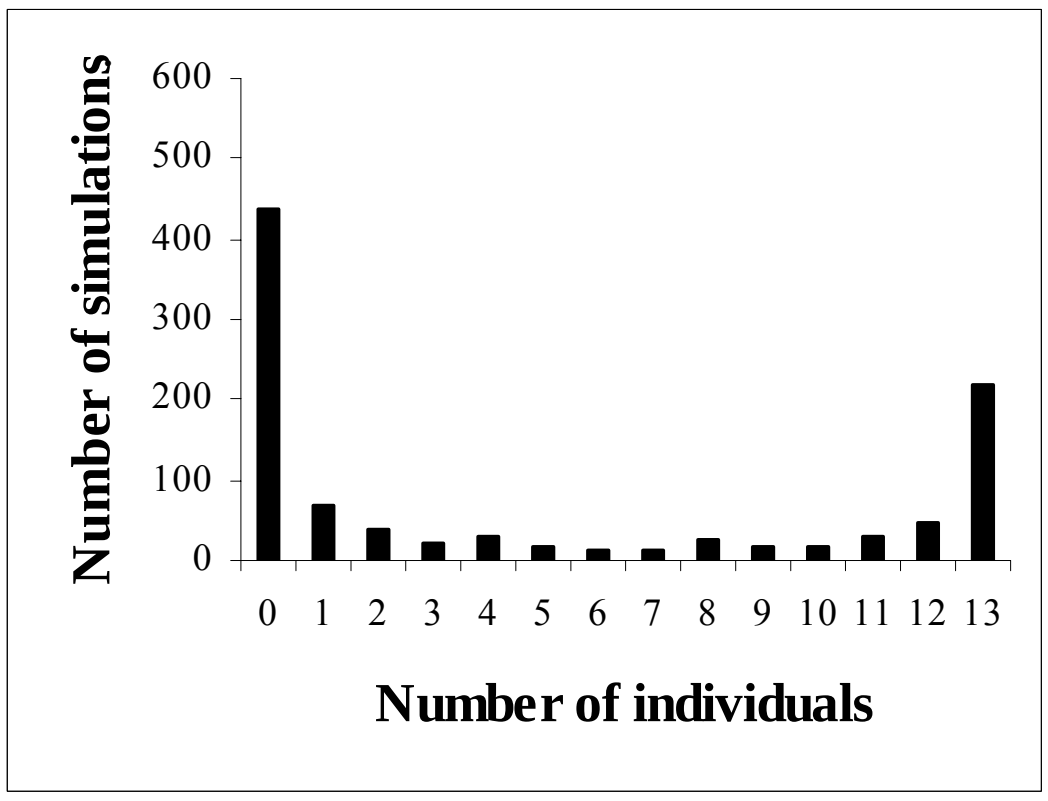

Number of simulations
600
500
400
300
200
100
0
0 1 2 3 4 5 6 7 8 9 10 11 12 13
Number of individuals

b

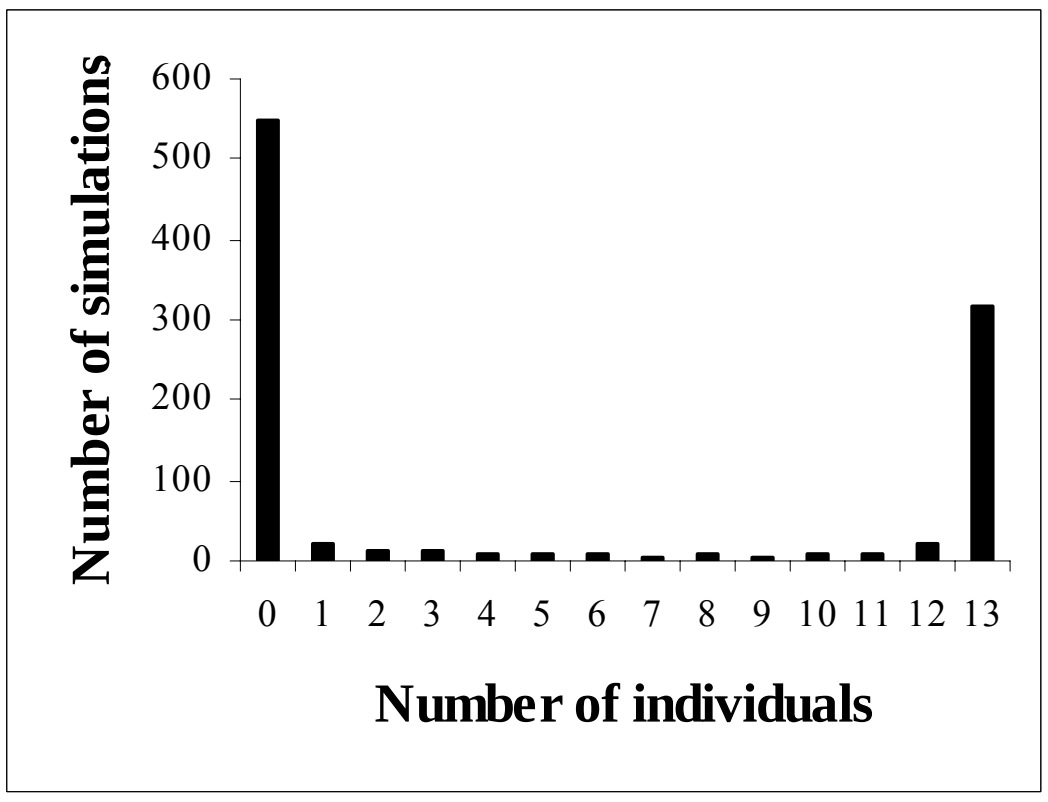

Number of simulations
600
500
400
300
200
100
0
0 1 2 3 4 5 6 7 8 9 10 11 12 13
Number of individuals

c

Figure 3

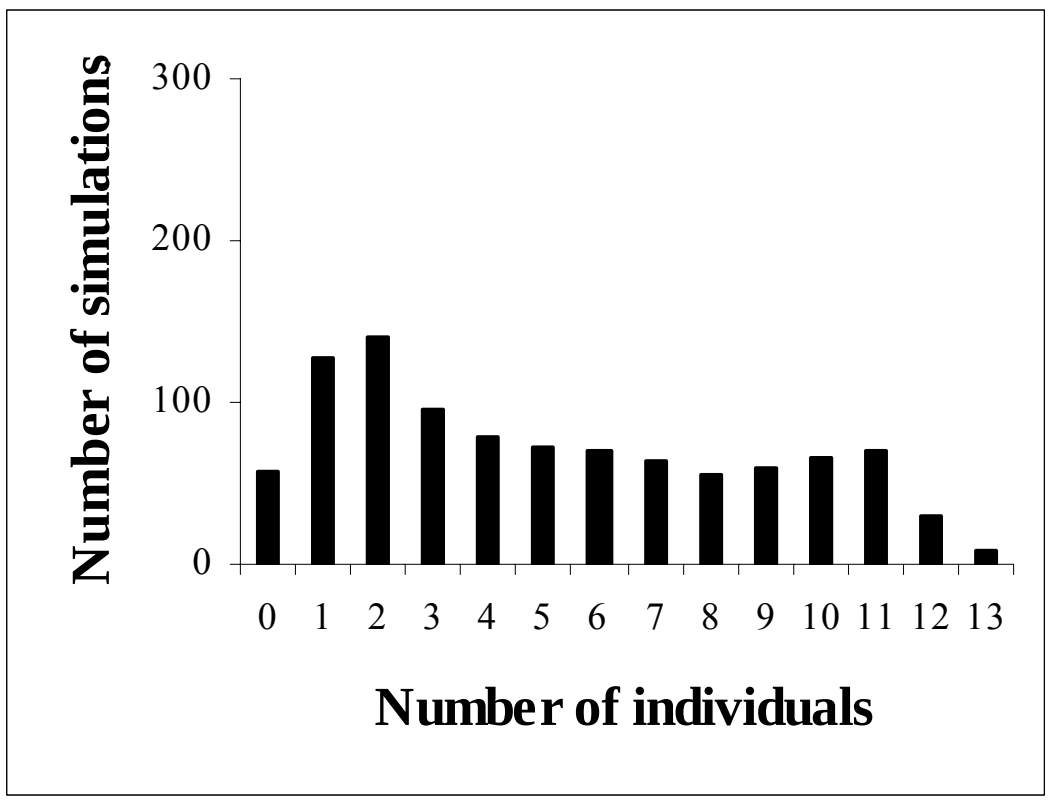

Number of simulations
300
200
100
0
0 1 2 3 4 5 6 7 8 9 10 11 12 13
Number of individuals

a

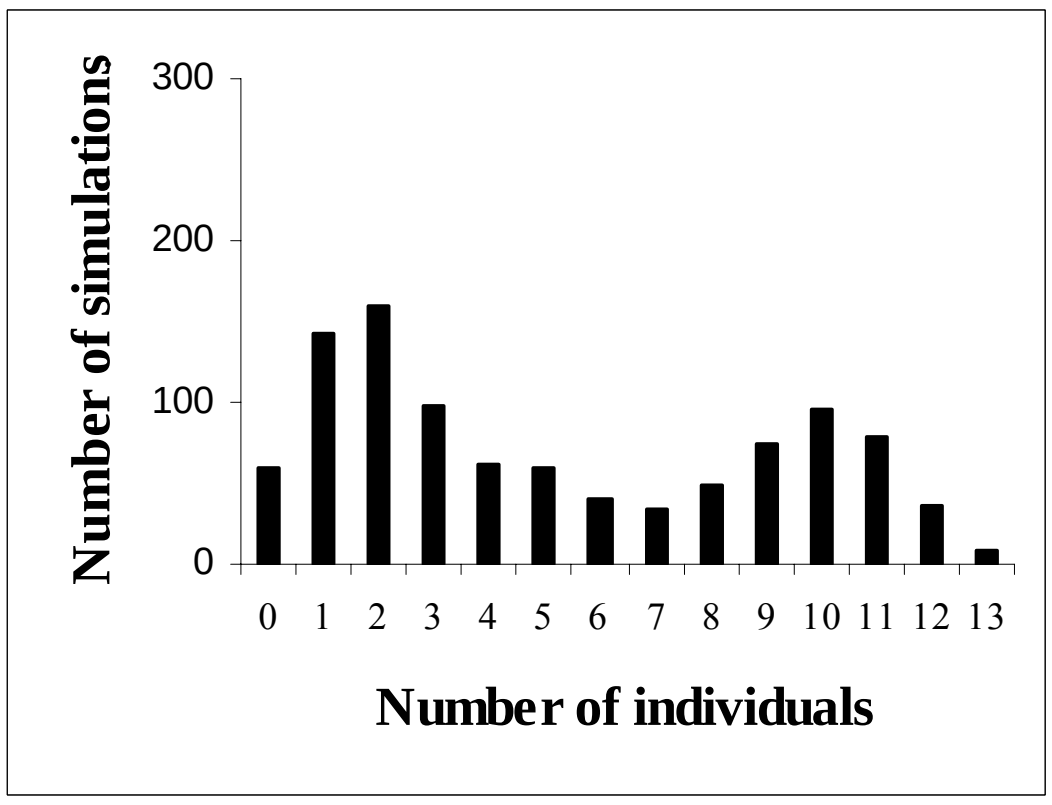

Number of simulations
300
200
100
0
0 1 2 3 4 5 6 7 8 9 10 11 12 13
Number of individuals

b

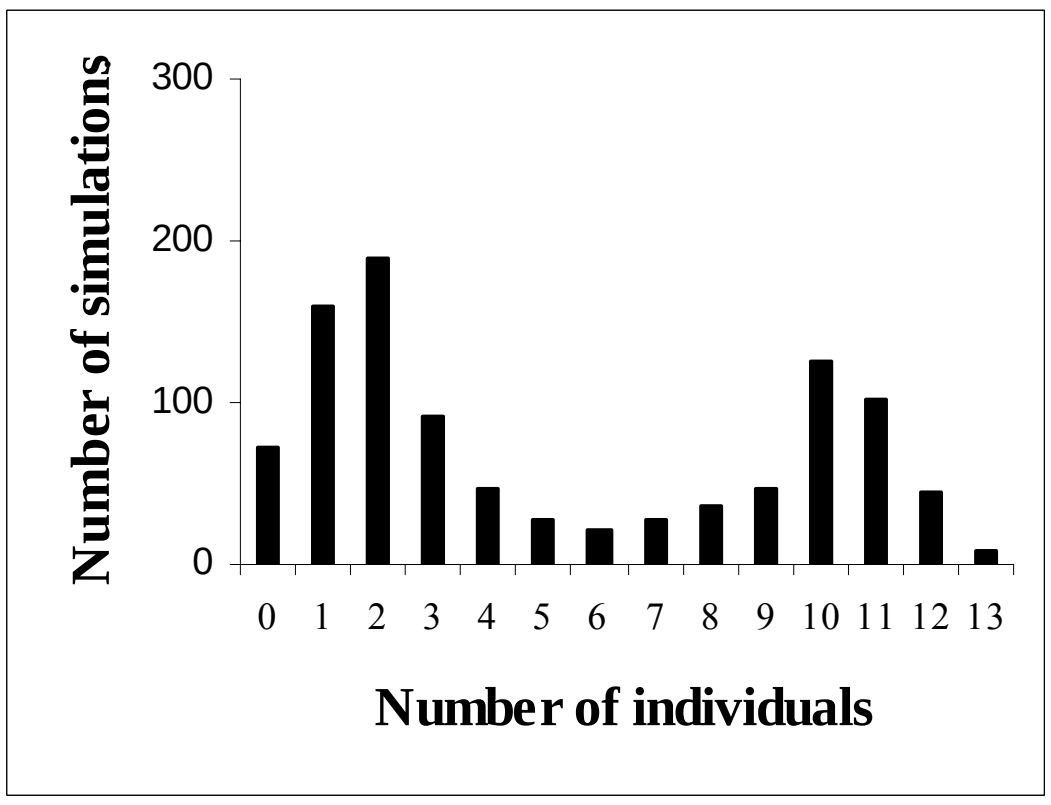

Number of simulations
300
200
100
0
0 1 2 3 4 5 6 7 8 9 10 11 12 13
Number of individuals

c

## Legends for figures

Figure 1

Theoretical distribution (a) and experimental distribution (b) of the number of tests where *i* (*i* = 0,...,13) individuals have chosen the left side. The theoretical distribution simulates the choices of individuals behaving independently.

Figure 2

Anonymous model: Individuals influenced by the animals which have already gone: Distribution of the simulations (1000 replications) as a function of the number of individuals having chosen the left side with three different mimetic coefficients $\beta$ = 1 (a), 3 (b) and 10 (c).

Figure 3

Individualised model: Distribution of the simulations (1000 replications) as a function of the number of individuals having chosen the left side with three different mimetic coefficients $\beta$ = 1 (a), 3 (b) and 10 (c).